\documentclass[prb,preprint,showpacs,floatfix,amsmath,longbibliography]{revtex4-1}

\usepackage{color}
\usepackage{dsfont}
\usepackage[normalem]{ulem}
\usepackage{dcolumn}
\usepackage[colorlinks=true,linkcolor=blue,citecolor=blue,urlcolor=blue]{hyperref}
\usepackage[titletoc,title]{appendix}

\usepackage{graphicx}
\def\be{\begin{equation}}
\def\ee{\end{equation}}
\def\beq{\begin{eqnarray}}
\def\eeq{\end{eqnarray}}

\newcommand{\unitmatr}{\ensuremath{\mathds{1}}}

\newcommand{\dd}{\ensuremath{\mathrm{d}}}

\def\doubleunderline#1{\underline{\underline{#1}}}

\begin{document}
\renewcommand{\familydefault}{\sfdefault}
\renewcommand{\sfdefault}{cmbr}

\title{Chiral magnetism of magnetic adatoms generated by Rashba electrons}
\author{Juba Bouaziz$^{1}$} 
\email{j.bouaziz@fz-juelich.de}
\author{Manuel dos Santos Dias$^1$}
\author{Abdelhamid Ziane$^{1,2}$}
\author{Mouloud Benakki$^2$}
\author{Stefan Bl\"{u}gel$^1$}
\author{Samir Lounis$^1$}
\email{s.lounis@fz-juelich.de}

\affiliation{$^1$ Peter Gr\"{u}nberg Institut and Institute for Advanced Simulation, 
Forschungszentrum J\"{u}lich and JARA, D-52425 J\"{u}lich, Germany}

\affiliation{$^2$ Laboratoire de Physique et Chimie Quantique, 
Facult\'{e} des Sciences, Universit\'{e} Mouloud Mammeri, 15000 Tizi-Ouzou, Algeria}

\begin{abstract}
We investigate long-range chiral magnetic interactions among adatoms mediated by 
surface states spin-splitted 
by spin-orbit coupling. Using the Rashba model, the tensor of exchange interactions 
is extracted wherein a the pseudo-dipolar interaction is found besides the usual isotropic 
exchange interaction and the Dzyaloshinskii-Moriya interaction. We find that, despite the latter interaction,
collinear magnetic states can still be stabilized by the pseudo-dipolar interaction. The inter-adatom 
distance controls the strength of these terms, which we exploit to design chiral magnetism 
in Fe nanostructures deposited on Au(111) surface. We demonstrate that these magnetic interactions are  
related to  superpositions of the out-of-plane and  in-plane components of the skyrmionic 
magnetic waves induced by the adatoms in the surrounding electron gas. We show that, even if the 
inter-atomic distance is large, the size and shape of the nanostructures dramatically impacts 
on the strength of the magnetic interactions, thereby affecting the magnetic ground state. We also
 derive an appealing connection between the isotropic exchange interaction and the 
Dzyaloshinskii-Moriya interaction, which relates the latter to the first order change of 
the former with respect to the spin-orbit coupling. This implies that the chirality defined 
by the direction of the Dzyaloshinskii-Moriya vector is driven by the variation of the isotropic 
exchange interaction due to the spin-orbit interaction.

\end{abstract}

\maketitle
\date{\today}


\section{Introduction}

The lack of inversion symmetry paired with strong spin-orbit (SO) coupling generate the 
Dzyaloshinskii-Moriya (DM) interaction~\cite{Dzyaloshinskii,Moriya}, a key ingredient for non-collinear magnetism, 
which is at the heart of chiral magnetism. 
The DM  interaction defines the rotation sense of the magnetization, rotating clockwise or counterclockwise along 
a given axis of a magnetic material. This is the case of spin-spirals in two-dimensional~\cite{Bode,Ferriani,Santos} or 
one-dimensional systems~\cite{Schweflinghaus,Menzel} down to   
zero-dimensional non-collinear metallic magnets~\cite{Khajetoorians,Mankovsky,Antal}. This type of interactions is 
decisive in the formation of the recently discovered magnetic 
skyrmions (see e.g. Refs.~\onlinecite{Muehlbauer,Yu,Romming,Heinze}), 
a particular class of chiral spin-texture, which existence was predicted three decades ago~\cite{Bogdanov,Roessler}. 
These structures are believed to be interesting candidates for future information 
technology\cite{Fert,Crum,Hanneken,Hamamoto} since 
lower currents are required for their manipulation, in comparison to conventional domain walls~\cite{Jonietz,Parkin}.

The ever-increasing interest in understanding the properties of the DM interaction and the 
corresponding vector is, thus, not surprising. Although the symmetry aspects of these interactions were discussed in 
the seminal work of Moriya~\cite{Moriya}, the ingredients affecting 
the magnitude and the particular orientation of a DM vector are not that explored 
but are certainly related to the details of the electronic structure. 
In the context of long-range interactions mediated by conduction electrons, 
the DM interaction was addressed by Smith\cite{Smith} and Fert and Levy\cite{Levy}. They found a strong analogy with 
the Ruderman-Kittel-Kasuya-Yosida (RKKY) interactions\cite{Ruderman,Kasuya,Yosida}. 
Indeed, the long-range DM vector oscillates in magnitude and changes 
its orientation as function 
of distance, which was recently confirmed experimentally with scanning tunneling microscopy (STM) and theoretically with 
ab-initio simulations based on density functional theory~\cite{Khajetoorians}. We note that today, besides theory, 
state-of-the-art STM experiments can be used to learn about the magnitude, oscillatory behavior and decay 
of the RKKY interactions as demonstrated in Refs.~\onlinecite{Zhou,Khajetoorians2,Prueser}.

Our goal is to address the DM interaction in an analytically tractable model and investigate 
its magnitude, sign and direction following 
a bottom-up approach assembling nanostructures of different sizes and shapes, atom-by-atom. 
We are particularly interested in the long-range magnetic interactions that have been already investigated several times 
theoretically. For example, Imamura et al.\cite{Bruno2004} considered pairs of localized spins interacting via the so-called 
two-dimensional Rashba gas of electrons~\cite{Rashba,Bychkov}  while Zhu et al.\cite{Zhu} replaced the Rashba gas by the surface of a 
topological insulator. We revisit the case of Rashba electrons and consider particularly 
the surface state of Au(111), where the Rashba spin splitting was observed experimentally\cite{Lashell}. We report on 
 selected nanostructures: dimers, wires, trimers, and two hexagonal structures deposited on the Au(111), 
where the interactions are mediated 
solely by the surface state. For the dimer case, we extract the analytical form 
of the magnetic exchange interactions tensor using the approximation of Imamura et al.\cite{Bruno2004}, labeled in the following 
RKKY-approximation, 
without renormalizing the electronic structure of the Rashba electrons because of the presence of the nanostructures. 
We found an inconsistency in the forms derived in Ref.~\onlinecite{Bruno2004}, a neglected integrable singularity observed at the minima of 
the energy dispersion curve, that we correct in the present article. Interestingly, we 
demonstrate that the magnetic interactions are intimately linked with 
the magnetization induced by the adatoms forming the dimers. We know, for instance, that a single magnetic adatom generates 
non-collinear magnetic Friedel oscillations, which can be decomposed into a linear combination of skyrmion-like magnetic 
waves~\cite{Lounis2012}. The in-plane components of the induced magnetization define the DM interaction, while the out-of-plane 
component is related to the usual RKKY-interaction. Also, we go beyond the RKKY-approximation by taking into account the impact of the deposited 
adatoms, which renormalize the electronic properties and can dramatically modify the long-range magnetic interactions. 
Moreover, we find an pseudo-dipolar term, or a two-ion anisotropy term generated by the presence of SO coupling, which 
plays a crucial role in the magnetism of the nanostructures. Although not carefully studied in the literature, these interactions can 
reach a large magnitude and counter-act the effect of the DM interaction by  
favoring collinear magnetism. After obtaining all magnetic interactions of interest, we use an extended Heisenberg model to 
investigate the magnetic states of the selected nanostructures.

\section{Description of the model}
The investigation of the magnetic behavior of the nanostructures is based on an 
embedding technique, where magnetic impurities are embedded on a surface characterized by the Rashba spin-splitted surface 
states. Once the electronic structure is obtained, we extract the tensor of magnetic exchange interactions as given in an 
extended Heisenberg model utilizing a mapping procedure described below. 

\subsection{Rashba model and embedding technique}
The twofold degenerate eigenstates of a two-dimensional electron gas confined in a surface 
or an interface, i.e.~a structure-asymmetric environment, experiences a spin-splitting induced by 
the spin-orbit interaction. 
 Within the model of Bychkov and Rashba\cite{Rashba,Bychkov}, this splitting effect is grasped by 
the so-called Rashba Hamiltonian 
\begin{equation}
\label{eq:Rashba_Ham}
\mathbf{H} = \frac{ p^2_{x} + p^2_{y} }{2m^*}~{\unitmatr}_2 - \frac{\alpha_\mathrm{so}}
{\hbar}(\boldsymbol{\sigma}_{x} {p_{y}} - \boldsymbol{\sigma}_{y} {p_{x}})\quad ,
\end{equation}
where ${p_{\gamma}}$, $\gamma \in \{x,y\}$, are the components of the momentum
operator $\vec{p}$ in a Cartesian coordinate system with $x,y$ coordinates in the surface plane whose surface normal 
points along $\hat e_z$.  $m^*$ is the effective mass of the electron. 
$\boldsymbol{\sigma}_\gamma$ are the Pauli matrices and $\unitmatr_2$ is the unit matrix in spin-space,
with the z-axis of the global spin frame of reference is parallel to $\hat e_z$. 
$\alpha_\mathrm{so}$ is the Rashba parameter,  a measure of the strength of the SO
interaction and the parameter that controls the degree of Rashba spin splitting.

The energy dispersion of the Rashba electrons is characterized by the $k$-linear splitting 
of the free-electron parabolic band dispersion:
\begin{equation}
E_{\mathrm{1(2)}} = \frac{\hbar^2}{2m^*}(k^2_{\mathrm{1(2)}} - k^2_{\mathrm{so}})\quad ,
\end{equation}
with $k_{\mathrm{1(2)}} = \sqrt{k_\mathrm{so}^2+\frac{2m^*E}{\hbar^2}} + (-)k_\mathrm{so}$ and $k_\mathrm{so} = \frac{m^* \alpha}{\hbar^2}$.
For the case of the surface state of the Au(111) surface, $\alpha=-0.4$ eV\,\AA~ and $m^{*} = 0.26\ m_{\mathrm{e}}$\cite{Walls}.
We want to calculate the magnetic interactions between magnetic adatoms deposited on a the Rashba electron gas. Therefore, we use an 
embedding technique, where we connect the Rashba Green function $\boldsymbol{G}^{0}$ to the Green function $\boldsymbol{G}$ of the system 
Rashba electron gas and magnetic adatoms via a Dyson equation. $\boldsymbol{G}^{0}$, connecting two points separated by $\vec{R}$, is given by:
\begin{equation}
\boldsymbol{G}^{0}(\vec{R},E+i\epsilon)=
\left(
\begin{array}{cc}
G_{\text{D}} &-G_{\text{ND}}\,e^{-i\beta}\\
G_\text{ND}\,e^{i\beta}&G_\text{D}\\
\end{array}
\right),
\end{equation}
where $G_D$ and $G_{ND}$, as defined in the appendix~\ref{appendixA}, depend on the position $\vec{R}$ and energy $E$, while $\beta$
is the angle between $\vec{R}$ and the $x$-axis. When magnetic adatoms are present, the Green function 
connecting the adatoms sites $i$ and $j$ can be obtained from the Dyson equation:
\begin{equation}
\boldsymbol{G}_{ij}(E) = \boldsymbol{G}_{ij}^{0}(E) + \sum_{km} \boldsymbol{G}_{ik}^{0}(E)\,\boldsymbol{T}_{km}(E)\,\boldsymbol{G}_{mj}^{0}(E)\quad ,
\label{renormalization}
\end{equation} 
Here $\boldsymbol{G}_{ij}^{0}(E)$ is the Rashba Green function connecting sites $i$ and $j$.
The full scattering matrix $\boldsymbol{T}(E)$ is given by a Dyson equation:
\begin{equation}
\boldsymbol{{T}}^{-1}_{ij}(E) = \boldsymbol{{t}}_{i}^{-1}(E)\,\delta_{ij} - \boldsymbol{G}^{0}_{ij}(E)\quad ,
\label{T-matrix2}
\end{equation}
Where $\boldsymbol{{t}}_{i}(E)$ is the single-site scattering matrix connected to the potential of  a single adatom 
$\boldsymbol{v}_{i}$ via:
 \begin{equation}
\boldsymbol{{t}}_{i}(E) = \boldsymbol{{v}}_{i} + \boldsymbol{v}_{i}\,\boldsymbol{G}^{0}_{ii}(E)\,\boldsymbol{{t}}_{i}(E)\quad,
\label{T-matrix}
\end{equation}
In practice, we proceed to the s-wave approximation~\cite{Heller,Lounis2012} since 
 the wavelength of Au(111) surface states at the Fermi energy and below are much larger than the size of a single adatom. 
In this approach, one can work with a single phase shift, 
$\delta_j(E)$, describing the scattering of the surface state at a single impurity: $\mathbf{t}_j= 
\frac{i\hbar^{2}}{m^{*}}\,(e^{2i\delta_j(E)}-1)$ at site $j$. 
\subsection{Extended Heisenberg model}
In the extended Heisenberg Hamiltonian $H$ given in , 
the elements of the magnetic exchange tensor, $\doubleunderline{{\mathbf{J}}}_{ij}$, can be extracted by  
differentiating $H$ according to $\vec{e}_{i}$ and $\vec{e}_{j}$: 
\begin{equation}
H = \sum_{i,j}\vec{e}_{i}\,\doubleunderline{{\mathbf{J}}}_{ij}\,\vec{e}_{j}\quad, 
J_{ij}^{\alpha \beta} = \frac{\partial^2 {H}}{\partial e_{i}^\alpha \partial e_{j}^\beta}\quad ,
\label{heis}
\end{equation}
with $\{\alpha, \beta\}=\{x,y,z\}$ and $\vec{e}_i$ being the unit vector 
of the magnetic moment at site $i$.
The exchange tensor is decomposed into three contributions:
\begin{equation}
\doubleunderline{{\mathbf{J}}}_{ij} = \frac{1}{3}\,\mathrm{Tr}\,\{ {\doubleunderline{\mathbf{J}}}_{ij}\}\,{\unitmatr}_{3}
+ {\doubleunderline{\mathbf{J}}}_{ij}^\text{A} + \doubleunderline{{\mathbf{J}}}_{ij}^\text{S}\quad .
\label{J+JA+JS}
\end{equation}
In the right-hand side of the previous equation, the first term is the isotropic exchange, while
$\doubleunderline{{\mathbf{J}}}_{ij}^\text{A}$ is the anti-symmetric part:
\begin{equation}
\doubleunderline{{\mathbf{J}}}_{ij}^\text{A} = \frac{\doubleunderline{{\mathbf{J}}}_{ij} - \doubleunderline{{\mathbf{J}}}_{ij}^\text{T}}{2}\quad ,
\end{equation}
it is connected to the Dzyaloshinskii-Moriya vector components via:
\begin{equation}
\doubleunderline{{\mathbf{J}}}_{ij}^\text{A} =
\left(
\begin{array}{rcl}
\ 0 &\ D_{ij}^z&\ -D_{ij}^y\\
\--D_{ij}^z &\ 0 &\ D_{ij}^x\\
\ D_{ij}^y &\ -D_{ij}^x &\ 0 
\end{array}
\right)\quad .
\end{equation}
The last term of Eq.~\ref{J+JA+JS}, $\doubleunderline{{\mathbf{J}}}_{ij}^\text{S}$, 
is the symmetric part which contains pseudo-dipolar interactions:
\begin{equation}
\doubleunderline{{\mathbf{J}}}_{ij}^\text{S} = \frac{\doubleunderline{{\mathbf{J}}}_{ij} +
 \doubleunderline{{\mathbf{J}}}_{ij}^\text{T}}{2} - \frac{1}{3} \, \mathrm{Tr} 
\, \{{\doubleunderline{\mathbf{J}}}_{ij}\}\,{\unitmatr}_{3}\quad . 
\end{equation}

\subsection{Mapping procedure}
The strategy is to consider the Hamiltonian describing the electronic structure of the nanostructures and perform 
the same type of differentiation as in Eq.~\ref{heis} in order to identify the tensor of magnetic exchange interactions. 
We use Lloyd's formula.~\cite{Lloyd1972}, 
which permits the evaluation of the energy variation due to an infinitesimal rotation of the 
magnetic moments, starting from a collinear configuration~\cite{Liech1987,udvardi2003,ebert2009}. 
In general, the contribution to the single-particle energy (band energy) after embedding the nanostructure is given by:
\begin{eqnarray}
E_{sp} =  \frac{1}{\pi} \, \mathrm{Im} \int_{E_R}^{E_{F}} \hspace{-3mm} \dd E \, \mathrm{Tr} 
\, \{ \mathrm{ln} \, \boldsymbol{T}(E)^{-1}\}\quad ,
\label{single-particle-energy}
\end{eqnarray}
where $E_F$ is the Fermi energy, $E_R = -\frac{\hbar^{2}k^2_{\mathrm{so}}}{2m^{*}}$ is the bottom of the Rashba electrons
 energy dispersion curve and Tr is the trace over impurity position- and 
spin-indices. 
The elements of the tensor of exchange interaction are then given by
\begin{equation}
J_{ij}^{\alpha \beta} =  \frac{\partial^2}{ \partial e_i^\alpha \partial e_j^\beta }E_{sp} = 
- \frac{1}{\pi} \, \mathrm{Im} \, \int_{E_{R}}^{E_{F}} \hspace{-3mm} \dd E
 \, \mathrm{Tr}\ \{\frac{ \, \partial^2}
{\partial e_{i}^\alpha \partial e_{j}^\beta}\,\mathrm{ln}\,\boldsymbol{T}(E)\}\quad .
\end{equation}
Using Eq.~\ref{T-matrix2}, we evaluate the required second derivative and find for the elements of the tensor of 
exchange interactions:
\begin{equation}
J_{ij}^{\alpha \beta}= -\frac{1}{\pi} \,\mathrm{Im} \, \int_{E_{R}}^{E_{F}} \hspace{-3mm} \dd E
 \,\mathrm{Tr}\, \{\mathbf{t}_i^\alpha \, \mathbf{G}_{ij} 
\, \mathbf{t}_j^\beta \, \mathbf{G}_{ji}\}\quad ,
\label{app}
\end{equation}
the trace is taken over the spin-index, and $\mathbf{t}_i^\alpha$ is 
simply the derivative of $\mathbf{t}$ with respect to $e_i^\alpha$. Since the $\mathbf{t}$-matrix can be written as:
\begin{equation}
\mathbf{t}= \frac{t_\uparrow+t_\downarrow}{2}\,
\unitmatr_2 
+
\frac{t_\uparrow-t_\downarrow}{2}
\,\vec{\boldsymbol{\sigma}} \cdot \vec{e}\quad ,
\label{t-matrix}
\end{equation}
we find that $\mathbf{t}_i^\alpha = \frac{\partial \mathbf{t}}{\partial e_i^{\alpha}} = \Delta_i\ {\boldsymbol{\sigma}}^{\alpha}$,
with $\Delta = \frac{t_\uparrow-t_\downarrow}{2}$. The final form of the tensor of magnetic 
exchange interactions is then finally given by:
\begin{equation}
J_{ij}^{\alpha \beta}= -\frac{1}{\pi}\, \mathrm{Im} \,  
\int_{E_R}^{E_F}\hspace{-3mm} \dd E\,\Delta_i\, \Delta_j\, \mathrm{Tr}\, \{ \boldsymbol{\sigma}^\alpha\, \mathbf{G}_{ij}\
 \boldsymbol{\sigma}^\beta \, \mathbf{G}_{ji}\}\quad .
\label{app2}
\end{equation}

\section{Magnetic properties of dimers}

\subsection{RKKY-approximation}
Before the numerical evaluation of the exchange tensor in nanostructures from Eq.~\ref{app2}, 
it would be interesting to have an approximated analytic form. This is achievable by considering in Eq.~\ref{app2}
the unrenormalized Green functions, $G_{0}$, instead of $G$. Here we recover the RKKY-approximation, 
expected from second order perturbation theory and used for example by Ref.~\onlinecite{Bruno2004}. 
In the particular case of a two-dimensional Rashba electron gas, the Rashba Green function can be 
expressed using Pauli matrices:
\begin{equation}
{\boldsymbol{G}}_{ij}^{0} = G_{D}\,\boldsymbol{\sigma}_{0} - i\,G_{ND}\,(\cos\beta\,\boldsymbol{\sigma}_y 
- \sin\beta\, \boldsymbol{\sigma}_x)\quad .
\label{G_pauli}
\end{equation}
Surprisingly, we found anisotropies in the diagonal part of the exchange
 tensor that are generally neglected in the literature.
The physical meaning of these anisotropies can be traced back to the extended 
Heisenberg model defined by the tensor of magnetic exchange interactions. 
In fact, by defining the $x$-axis as the line connecting the two sites 
$i$ and $j$, we show in appendix~\ref{appendixB} that the extended Heisenberg 
Hamiltonian describing the corresponding magnetic coupling 
can be written as: 
\begin{equation}
\vec{e}_{i}\,\doubleunderline{{\mathbf{J}}}_{ij}\,\vec{e}_j= J\,\vec{e}_i\cdot\vec{e}_j + D\,(\vec{e}_i\times\vec{e}_j)_y + I\,{e}_i^y\,{e}_j^y\quad ,
\label{zest}
\end{equation}
where the exchange constants $\{ J, D$ and $I\}$ are related to the Rashba Green function by: 
\begin{equation}
J = - \frac{2}{\pi}\,\mathrm{Im}\,\int_{E_{R}}^{E_{F}}\hspace{-3mm} \dd E\,\Delta_i\,\Delta_j\,(G_{D}^2-G_{ND}^2)\quad ,
\label{ech1}
\end{equation}
\begin{equation}
D = \frac{4}{\pi}\,\mathrm{Im}\,\int_{E_{R}}^{E_{F}}\hspace{-3mm} \dd E\,\Delta_i\,\Delta_j\,G_{D}\,G_{ND}\quad ,
\label{ech2}
\end{equation}
\begin{equation}
I = -\frac{4}{\pi}\,\mathrm{Im}\,\int_{E_{R}}^{E_{F}}\hspace{-3mm} \dd E\,\Delta_i\,\Delta_j\,G_{ND}^2\quad .
\label{ech3}
\end{equation}
$J$ is the isotropic exchange interaction, which if positive favors an 
anti-ferromagnetic coupling in our convention, otherwise it favors a ferromagnetic coupling. 
$D$ is the $y$ component of the D vector, which is by symmetry the only nonzero component (Third rule of Moriya~\cite{Moriya}). 
This favors chiral magnetic textures lying in the $xz$ plane. $I$ is the pseudo-dipolar term, a two-ion anisotropy term, 
coming from the symmetric part of the exchange tensor. It leads to an anisotropy in the diagonal-part of the 
tensor of exchange interaction, for instance ${J}_{ij}^{xx} = {J}_{ij}^{zz} \neq {J}_{ij}^{yy}$. 
Considering the impurities along the $x$-axis, $I$ is given by ${J}_{ij}^{yy}-{J}_{ij}^{zz}$. 
This anisotropy is finite because of the two-dimensional nature of the Rashba electrons, so 
the $x$- and $y$-directions are nonequivalent to the $z$-direction. Here, $I$ 
favors a collinear magnetic structure along the $y$-axis and counteracts the DM interaction.
The analytical forms of the magnetic exchange interactions allow us to understand their origin in terms of the magnetic 
Friedel oscillations generated by single atoms.\cite{Lounis2012} These oscillations carry a complex magnetic texture 
that can be interpreted in terms of skyrmionic-like waves. 
Within the RKKY-approximation and neglecting the energy dependence of $\Delta_{i}$, the isotropic interaction, $J$, connecting two 
impurities at site $i$ and $j$, is proportional to the $z$-component of magnetization generated at site $j$ by a 
single impurity at site $i$. In other words, the impurity at site $j$ feels the effective magnetic field generated by the magnetization 
at that site but induced by the adatom at site $i$. $D$, however,  is defined by the in-plane component of the induced 
magnetization. This is a central result of our work. Here, the corresponding magnetic field felt by the second impurity has an in-plane component 
and naturally leads to a non-collinear magnetic behavior, {\it i.e.} the natural impact of the DM vector.  
$I$ does not have a simple interpretation, but it can be related to the anisotropy (difference) of the induced 
magnetization parallel to the impurity moment upon its rotation from out of plane to in plane.
In the following we proceed to
 the analytical evaluation of \{$J, D, I$\} from the equations above. The details of the integration are given in appendix~\ref{appendixC}.

{\bf{Evaluation of J.}} In order to derive analytically the exchange interactions, we use
 an approximation for the $\mathbf{t}$-matrices. We assume that they are energy independent 
(Resonant scattering for the minority-spin channel, {\it i.e.} $\delta_{\downarrow} = \frac{\pi}{2}$, and no scattering for the majority-spin channel,
{\it i.e.} $\delta_{\uparrow} = {\pi}$), 
which allows us to write $\Delta_i\,= -\frac{2i\hbar^{2}}{m^{*}}$ . This approximation used in Ref.~\onlinecite{Lounis2012} is reasonable 
for an adatom like Fe deposited on Au(111) surface. Then, we find the asymptotic behavior of $G_{D}$ and $G_{ND}$ for large 
distances $R$ (see appendix~\ref{appendixC}). The isotropic exchange constant can be expressed as:
\begin{equation}
\begin{split}
  J &= \frac{2}{\pi^2 R}\,\mathrm{Im}\int_{E_{R}}^{E_{F}}\hspace{-3mm}\dd E\,\frac{i}{(k_1+k_2)^2}\,(k_1\,e^{2ik_1R} + k_2\,e^{2ik_2R})\\
   &= \frac{\hbar^2}{m^*\pi^2 R}\Bigg{[}\,\frac{1}{2R}\sin(2k_{F}R)\cos(2k_\mathrm{so}R) - k_\mathrm{so}\sin(2k_\mathrm{so}R) 
\ \mathrm{SI}(2k_{F}R)\,\Bigg{]}\quad ,
\end{split}
\label{J_analytic}
\end{equation} 
where $\mathrm{SI}(x)$ is the sine-integrated function of $x$.  $J$  is found to be the sum of two 
functions. The first one evolves as a function of  $\frac{1}{R^2}$, as expected for regular two-dimensional 
systems but the second function decays like $\frac{1}{R}$, which 
has been neglected in the work of Ref.~\onlinecite{Bruno2004}. The $\frac{1}{R}$ decay leads to a slower 
decay of $J$ than what is known for a regular two-dimensional electron gas. It comes from the Van Hove
singularity at the bottom of the conduction band at energies below the crossing of the two $k$-splitted bands of the 
energy dispersion. At very larges distances, $\mathrm{SI}(x)$ converges to a constant ($\frac{\pi}{2}$) 
and $J$ is behaving like $\frac{1}{R}\sin(2k_\mathrm{so}R)$. Naturally, when $k_\mathrm{so}$ is 
set to zero we recover the classical form of the RKKY interaction without spin-orbit coupling for a free 
electron gas in two-dimensions, i.e.~$J$ evolves like $\frac{1}{R^2}\sin(2k_{F}R)$. 

{\bf{Evaluation of D.}} We consider the same approximations used above to calculate 
the $y$-component of the DM vector ($D$) 
and find: 
\begin{equation}
\begin{split}
D &=  -\frac{4}{\pi^2 R}\,\mathrm{Im}\int_{E_{R}}^{E_{F}}\hspace{-3mm} \dd E \,\frac{1}
{2(k_1+k_2)^2}\,[\,k_1\,e^{2ik_1R} -  k_2\,e^{2ik_2R}\,]\quad ,\\
&=  - \frac{\hbar^2}{m^*\pi^2 R}\,\Bigg{[}\,\frac{1}{2R}\sin(2k_{F}R)
\sin(2k_\mathrm{so}R)+ k_\mathrm{so}\,\cos(2k_\mathrm{so}R)\, 
\mathrm{SI}(2k_{F}R)\,\Bigg{]}\quad .
\end{split}
\label{D_analytic}
\end{equation} 
Like the isotropic exchange constant, $D$ is a sum of two terms. The first term decays 
as $\frac{1}{R^2}$ while the second as $\frac{1}{R}$. 
A perturbative development of $D$ in terms of $k_{so}$ shows that $D$ is first order in 
spin-orbit coupling. At very large distances $D$ evolves like $\frac{1}{R}\cos(2k_\mathrm{so}R)$.  

{\bf{Evaluation of I.}} In appendix~\ref{appendixC}, we show that $I$ is a sum of two integrals 
over the energy because of a branch cut in the Hankel functions.  
The first integral, denoted $I_1$, goes from $E_R$ to zero and the second, $I_2$, goes from  zero to $E_{F}$.
\begin{equation}
\begin{split}
I_{1}&= -\frac{4}{\pi^2 R}\,\mathrm{Im} \int_{E_{R}}^{0}\hspace{-3mm} \dd E\,\frac{1}{2(k_1+k_2)^2}\,\\
&[\,i\,(k_1\, e^{2ik_1R} + k_2\, e^{2ik_2R})
+2 \sqrt{|k_1| k_2}\, e^{i(k_2 - |k_1|)R}\,\quad ,\\
I_{2}&= -\frac{4}{\pi^2 R }\,\mathrm{Im} \int_{0}^{E_{F}}\hspace{-3mm} \dd E\,\frac{i}{2(k_1+k_2)^2} \\ 
& [\, k_1\,e^{2ik_1R} + k_2 \,e^{2ik_2R}\,-2\sqrt{k_1 k_2}\,e^{i(k_1 + k_2)R}\,]\quad ,
\end{split}
\end{equation}
and if we sum up the two terms:
\begin{equation}
I = -J + \frac{\hbar^2}{m^* \pi^2 R}\Bigg{[}  
   \int_{|k_\mathrm{so}|}^{k_{F}}\hspace{-3mm} \dd q\, \sqrt{1-\frac{k_\mathrm{so}^2}{q^2}}\,\cos(2qR)
-  \int_{0}^{|k_\mathrm{so}|}\hspace{-5mm} \dd q\,\sqrt{\frac{k_\mathrm{so}^2}{q^2}-1}\,\sin(2qR)\,\Bigg{]}\quad .
\label{I_analytic}
\end{equation} 
The integral involving $\cos(2qR)$ is important at short distances since it competes with one of the terms defining 
$-J$. In fact, it has the opposite sign of $-\frac{1}{2R}
\sin(2k_{F}R)\cos(2k_\mathrm{so}R)$ (see Eq.~\ref{J_analytic}). 
This reduces considerably the value of $I$ comparing to $J$. The second integral involves $\sin(2qR) $ and therefore it 
leads to a small contribution for low values of $k_\mathrm{so}$. A perturbative development of $I$ in terms of $k_\mathrm{so}$ shows that $I$ is 
second order in spin-orbit coupling $(\propto k_\mathrm{so}^2)$. 

\begin{figure}[ht]
  \centering
  \includegraphics[width= 5 in, angle=0]{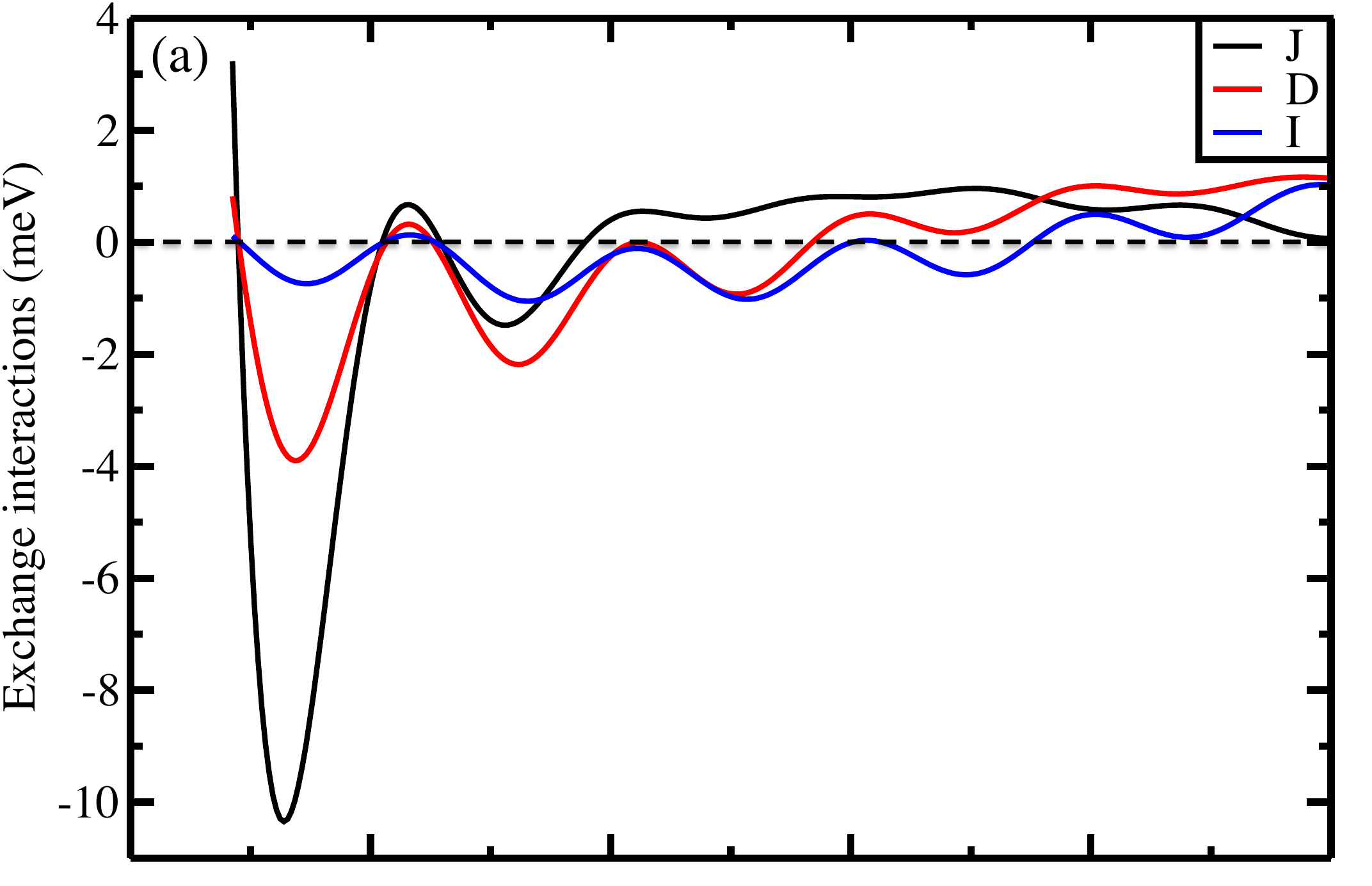}
  \includegraphics[width= 5 in, angle=0]{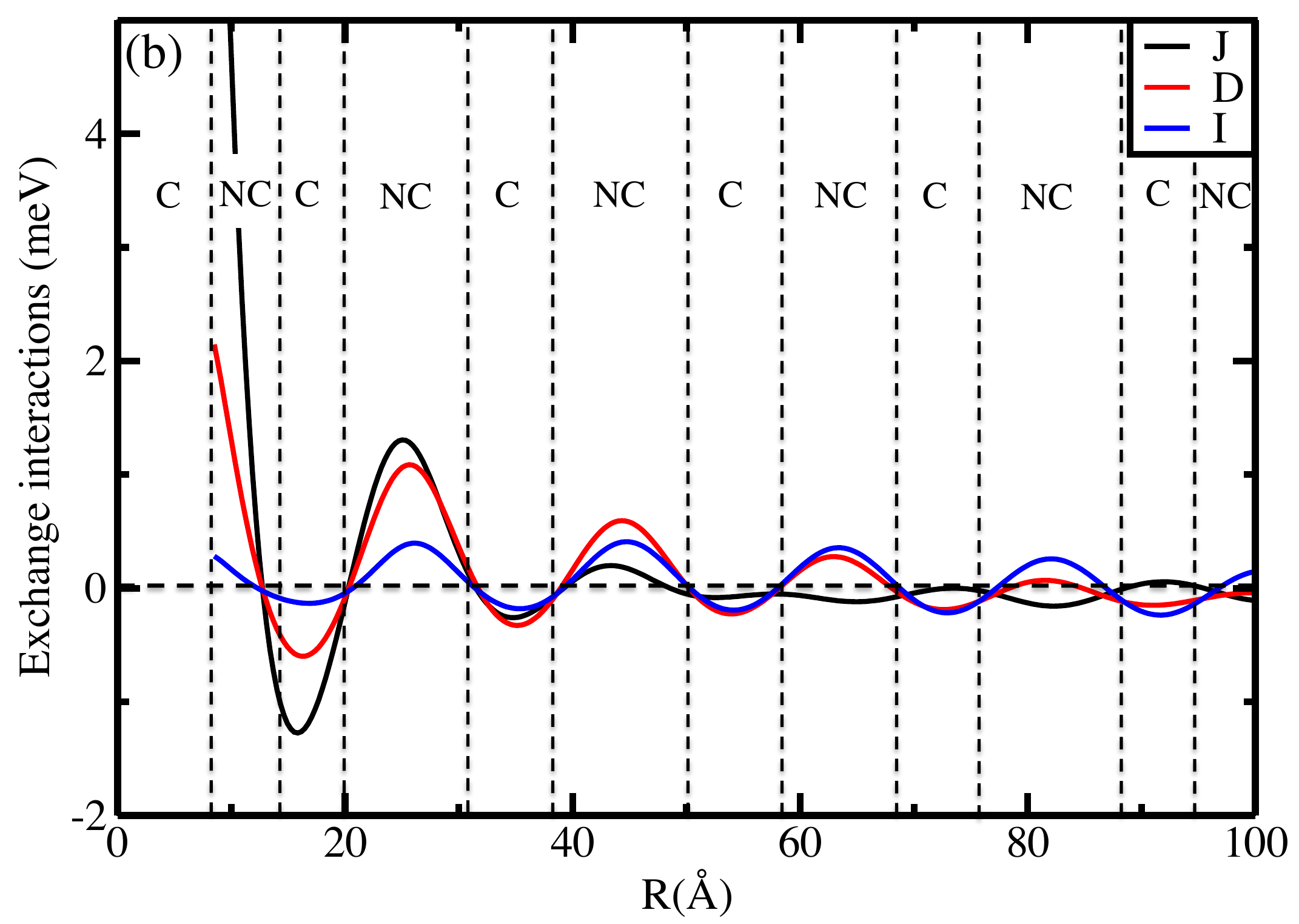}
  \caption{Evolution of the magnetic interactions $J$, $D$, $I$ as a function of the distance by using the RKKY approximation,
  for $\alpha=-0.4$ eV\,\AA~and $m^*=0.26\hspace{1mm} m_\mathrm{e}$ (parameters for Au(111) surface~\cite{Walls} used in Eq.~\ref{eq:Rashba_Ham}). (a) We use the 
RKKY-approximation (see Eqs.~\ref{ech1}, \ref{ech2}, \ref{ech3}) and assume a maximal scattering cross section for the minority spin 
channel ($\delta_\downarrow = \frac{\pi}{2}$) and no contribution for the majority spin channel ($\delta_\uparrow = \pi$). 
(b) We go beyond the RKKY-approximation and use the renormalized electronic structure (Eqs.~\ref{renormalization} and \ref{app2}) induced by the presence of two impurities. 
The vertical lines define a magnetic phase diagram indicating the nature of the orientation of the two magnetic moments 
as function of their mutual distance. C indicates the collinear phase of the magnetic moments and NC the non-collinear phase.}
  \label{echangeG0}
\end{figure}

In Fig.~\ref{echangeG0}(a), we plot the magnetic exchange interactions $J, D$ and $I$ 
as function of the distance between two magnetic adatoms. The black curve depicts $J$, which at short distances is 
characterized by a wavelength of $\frac{2\pi}{k_F} \sim 37\,$\AA. Indeed, from Eq.~\ref{J_analytic} we expect a beating of 
the oscillations when $\cos(2k_\mathrm{so}R) = 0$, in other words at $R = \frac{\pi}{4k_\mathrm{so}}\sim60\,$\AA.
The latter results from the SO interaction and it is connected to the SO wavelength, $\frac{2\pi}{k_\mathrm{so}}\sim 480\,$\AA. 
One notices that for a large range of distances ($R > 25$ \AA) the magnetic interactions do not oscillate around the $y = 0$ axis. 
This is an artifact of the RKKY-approximation. Similarly to $J$, $D$ is negative for distances larger than $25$ \AA, which means within
 the RKKY-approximation, the chirality defined by the sign of the DM interaction changes only for dimers separated by rather small distances. 
We notice also that $D$ and $I$ are oscillating functions that can be of the same magnitude as $J$. Thus, we believe that 
such systems provide the perfect playground to investigate large regions of the magnetic phase diagram inaccessible with 
usual magnetic materials.  

\subsection{Beyond the RKKY-approximation}
The deposited magnetic impurities naturally renormalize the electronic properties of the Rashba electrons. 
To evaluate their impact on the electronic states mediating the magnetic exchange interaction, 
we numerically compute $\boldsymbol{G}$, by considering consistently the multiple scattering effects. 
This is done first via considering an energy dependence in the $\mathbf{t}$-matrix assuming 
that they correspond to a Lorentzian in the electronic structure of the impurities and thus the phase shift 
is given by $\delta_{\sigma}(E) = \frac{\pi}{2} + \mathrm{atan}\bigg{(}{\frac{E-E_{\sigma}}{\Gamma}}\bigg{)}$. 
The parameters are extracted from ab-initio~\cite{Juba}, with a band width $\Gamma$ = $0.3\,$eV, and 
$E_{\downarrow} = 0.54\,$eV for the minority-spin channel, slightly on top of the Fermi level
 $E_{F} = 0.41\,$eV, the exchange splitting is $2.8\,$eV with respect to the majority spin-resonance 
($E_{\uparrow} = E_{\downarrow}  - 2.8\,$eV). Then we use (Eq.~\ref{T-matrix2}) for computing 
$\boldsymbol{T}(E)$. Afterwards we solve the Dyson equation (Eq.~\ref{renormalization}) giving $\boldsymbol{G}$.
The evolution of the three exchange interactions after renormalizing the Green function is given in Fig.~\ref{echangeG0}(b). 
Also we note the disappearance of the RKKY-approximation artifact leading to a non-change of sign of the exchange 
interaction at very large distances. Indeed,  contrary to the curves obtained with the RKKY-approximation, the magnetic interactions 
oscillate around zero. We traced back this effect to the strong reduction of the quasi one-dimensional behavior of the Rashba 
electron gas at $E_R$ because of the presence of the adatoms. This washes out the sine integrals seen for example 
in Eqs.~\ref{J_analytic} and \ref{D_analytic}.

The beating effect in $J$ occurs at the same distance as in the RKKY-approximation because it is an 
intrinsic property of the Rashba electron gas. 
At large distances the intensities of $J$ and $D$ are decreasing quickly, 
but $I$ keeps oscillating  up to a distance of $\sim 200$ \AA{} where it decreases quickly to zero. 

\subsection{Magnetic configurations of dimers} 
Having established the behavior of the tensor of magnetic exchange interactions as a function of distance, we investigate now
the magnetic ground state of different nanostructures characterized by different geometries and different sizes. 
After getting the magnetic interactions with the mapping procedure described above, we minimize the 
extended Heisenberg Hamiltonian with respect to the spherical angles, $(\theta_i,\phi_i)$, defining the orientation
of every magnetic moment $\vec{e}_i = (\cos\phi_i\sin\theta_i, \sin\phi_i\sin\theta_i, \cos\theta_i)$.
In order to check the stability of the magnetic ground state, we often add to 
the extended Heisenberg Hamiltonian the term $K \sum_{i}{e^{z}_{i}}^{\,2}$, where $K$ is a single-ion magnetic anisotropy energy favoring 
an out-of-plane orientation of the magnetic moment as it is the case for an Fe adatom on Au(111). We choose as a typical 
value $K = -6$~meV for all the investigated nanostructures~\cite{Spir}.

For the particular case of the dimer, an analytical solution is achievable by noticing that 
two magnetic states are possible: collinear (C) and non-collinear (NC). This is counter-intuitive, since the 
presence of the DM interaction leads usually to a non-collinear ground state. The presence of the pseudo-dipolar term $I$ makes 
the physics richer and stabilizes collinear magnetic states. 
Once more, because of the particular symmetry provided by the Rashba electron gas, within 
the non-collinear phase, the only finite component of the DM vector, $D_y$, enforces the two magnetic moments to lie in $xz$ plane 
perpendicular to the DM vector. Within the collinear phase, $I$ enforces the moments to point along the $y$-axis. 
 
{\bf{Non-collinear phase}.} Here the magnetic moments lie in the $xz$ plane and the pseudo-dipolar term
 does not contribute to the ground state energy. The ground state is then defined by
 the angle, $\theta_0 = \mathrm{atan}\frac{D}{J}$, between the two magnetic moments at sites $i$ and $j$. 
The energy corresponding to this state is $-|J|\sqrt{1 + \frac{D^2}{J^2}}$. 
With the single-ion anisotropy, $K$, the ground state angle becomes 
 $\theta_{0} = \hspace{1mm}\mathrm{atan}\hspace{1mm}(\frac{D}{J+K})$.  
As an example, we consider two adatoms separated by $d=10.42$~\AA~which corresponds
 to the seventh nearest neighbors distance on Au(111). In this case $J = 3.45$ meV and $D = 0.96$ meV and 
the ground state angle  ($\theta_{0}$) is $164^\circ$  $(K = 0\,\mathrm{meV})$ or $171^\circ$ $(K = -6\,\mathrm{meV})$. 

{\bf{Collinear phase}.} Here $D$ does not contribute to the ground state configuration. When $J$ and $I$ are both negative
the magnetic moments are parallel and point along the axis $y$-axis with the energy $J+I$, while for positive $J$ and $I$ 
the magnetic moments are anti-parallel 
and point along the $y$ axis too, with the energy $-(J+I)$. If $J$ and $I$ have opposite signs, for $J > 0$ the magnetic 
moments are anti-parallel in the $(xz)$--plane with the energy $-J$, while for $J < 0$ the magnetic moments are parallel in the
 $(xz)$--plane with the energy $J$. However, these last two solutions will not occur, since the NC phase is lower in energy.

There is competition between the collinear phase C and the non-collinear phase NC, which depends on the involved magnetic 
interactions. Without $I$, Fig.~\ref{echangeG0}(b) will consist of one single phase, the NC phase. Thanks to $I$, there is an 
alternation of the two phases depending on the inter-adatom distance. The magnetic anisotropy $K$ 
favors an out-of-plane orientation of the moments and tends to decrease the spatial range of the collinear 
phase where the moments point along the $y$-axis.

\begin{figure}[ht]
  \centering
  \includegraphics[width= 6 in, clip=true]{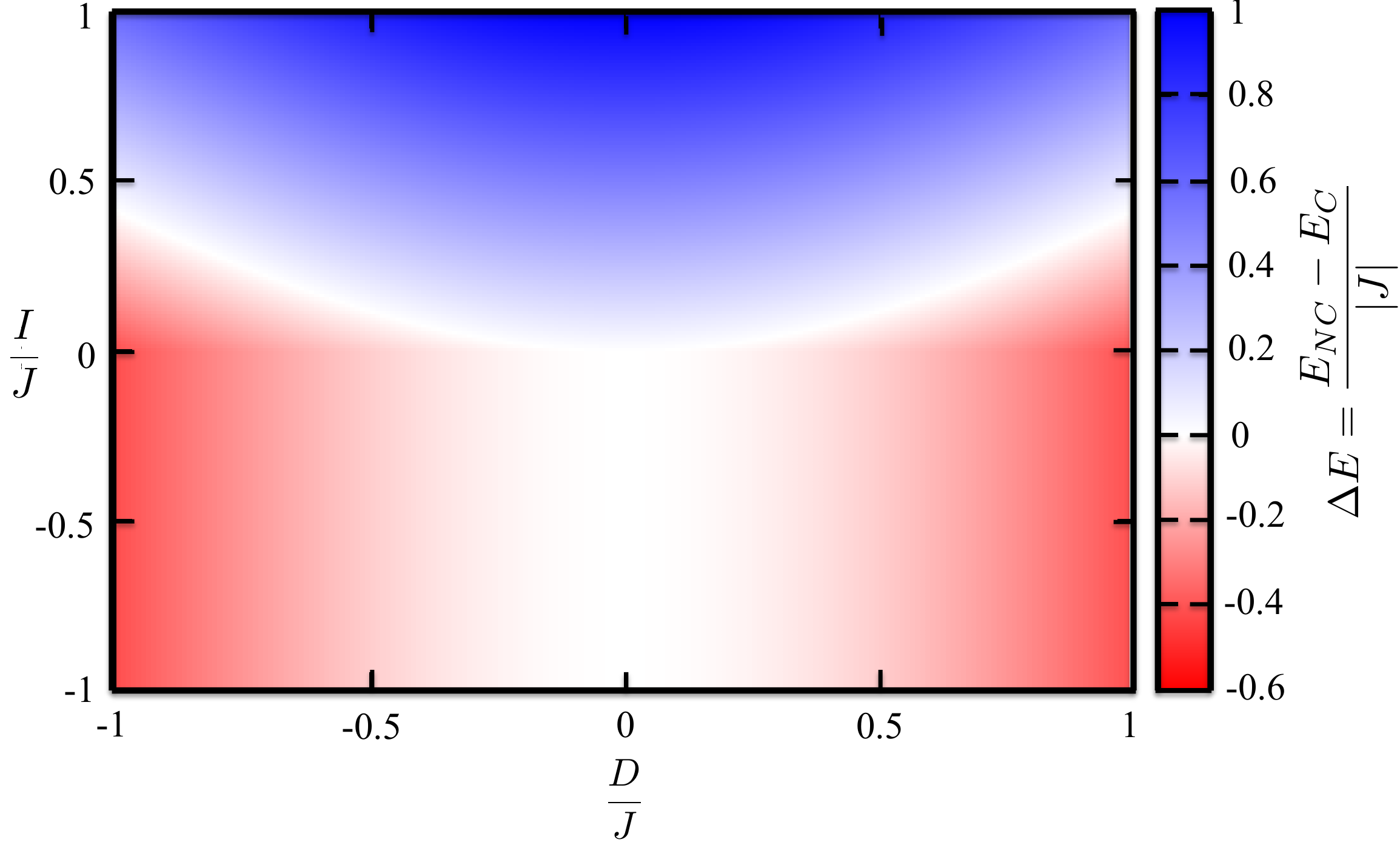}
  \caption{Phase diagram for the magnetic ground states of dimers. The color scale 
represents the energy difference normalized by $|J|$ between the non-collinear (red color) and collinear states (blue color) 
as function of the parameters \{$\frac{D}{J},\frac{I}{J}$\} (see Eq.~\ref{phase_diagram}).}
  \label{phdm}
\end{figure}

{\bf{Phase diagram}.} In Fig.~\ref{phdm}, we plot the phase diagram of the dimers $(K = 0\,\mathrm{meV})$. The color scale shows 
the energy difference ${\Delta E}$ between the ground states found in the NC phase and 
C phase normalized by $|J|$. A negative (positive) energy difference 
corresponds to a NC (C) ground state. Thus the blue region corresponds to a C phase and the red region to a NC phase:
\begin{equation}
\begin{cases}
&\Delta E = \frac{E_{NC} - E_{C}}{|J|} = -\sqrt{1 + \frac{D^2}{J^2}} +1\quad \text{for}\ J\ \text{and}\ I\ \text{with an opposite sign,} \\
&\Delta E = \frac{E_{NC} - E_{C}}{|J|} = -\sqrt{1 + \frac{D^2}{J^2}}+(1+\frac{I}{J})\quad \text{for}\ J\ \text{and}\ I\ \text{with the same sign.} \\
\end{cases}
\label{phase_diagram}
\end{equation}
For small ratios $\frac{D}{J}$,  if $\frac{I}{J} < 0$ then $\frac{\Delta E}{|J|}$ simplifies to 
$-\frac{D^2}{2J^2}$ and if $\frac{I}{J} > 0$ it simplifies to $-\frac{D^2}{2J^2} + \frac{I}{J}$, which define 
the magnetic phases plotted in Fig.~\ref{phdm}. We notice that when $I$ and $J$ are of the same 
sign, the dimers are mostly characterized by a C ground state. The corresponding C phase is separated from the NC phase by a parabola as 
expected from the term $-\frac{D^2}{2J^2}$. Moreover we note that even within the NC phase, a transition occurs 
when the sign of $\frac{I}{J}$ changes. This is related to the nature of the NC phase that changes by switching the sign of $\frac{I}{J}$, 
which leads to an additional, $\frac{I}{J}$, term in the energy difference. As mentioned earlier, if $\frac{I}{J}$ is positive the moments 
are in plane and align (parallel or anti-parallel) along the $y$ direction, while a negative $\frac{I}{J}$ leads to an alignment in the $(xz)$ plane. 
For negative $\frac{I}{J}$, one notices that when $\frac{D}{J}$ goes to zero, the plotted energy difference goes to zero, which does not mean that 
the C and NC phases are degenerate but it is the signature that the rotation angle of the moments goes to zero. Thus at $\frac{D}{J} = 0$ we have only 
a C phase.

{\bf Connecting $J$ to $D$.}  Before investigating nanostructures containing more than two adatoms, it is interesting 
to analyze the possibility of connecting $J$ to $D$. Recently, it was demonstrated that in the context of a micromagnetic model, 
the spin stiffness $A \sim \sum_j R_j^2 J(R_j)$, the micromagnetic counterpart of $J$, and $L = \sum_j R_j D(R_j)$, 
the counterpart of $D$ called the Lifshitz invariant can be related to each other for low SO interaction~\cite{Kyoung-Whan}:
\begin{equation}
L \sim  -2 k_\mathrm{so} A\quad .
\label{LA} 
\end{equation}
The sum over sites $j$ is limited by the size of the nanostructure but it can be infinite, e.g. if dealing with a monolayer 
or an infinite wire. 

We checked the validity of the previous relation utilizing the analytical forms of $J$ and $D$ obtained in the RKKY-approximation, i.e. Eqs.~\ref{J_analytic} 
and ~\ref{D_analytic},  and found that Eq.~\ref{LA} can be recovered for $k_{so}R << 1$
 but the error is proportional to the term involving the sine integral $\mathrm{SI}(2k_{F}R)$. So 
if one neglects the quasi one-dimensional behavior of the Rashba gas, one gets the formula of Kim et al.\cite{Kyoung-Whan}. 

\begin{figure}[ht]
  \centering
  \includegraphics[clip=true, width=5 in, angle=0]{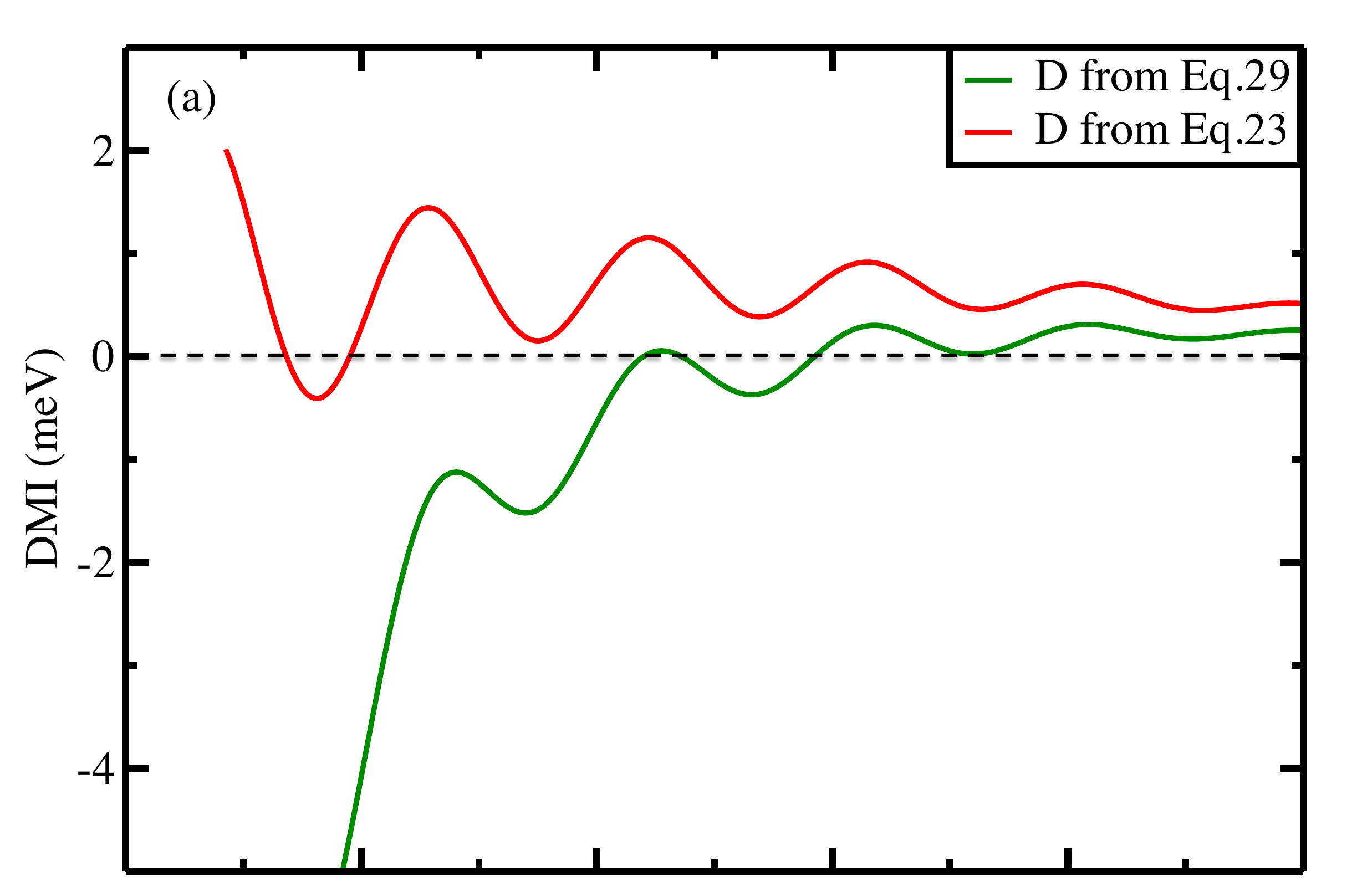}
  \includegraphics[width= 5 in,  angle=0]{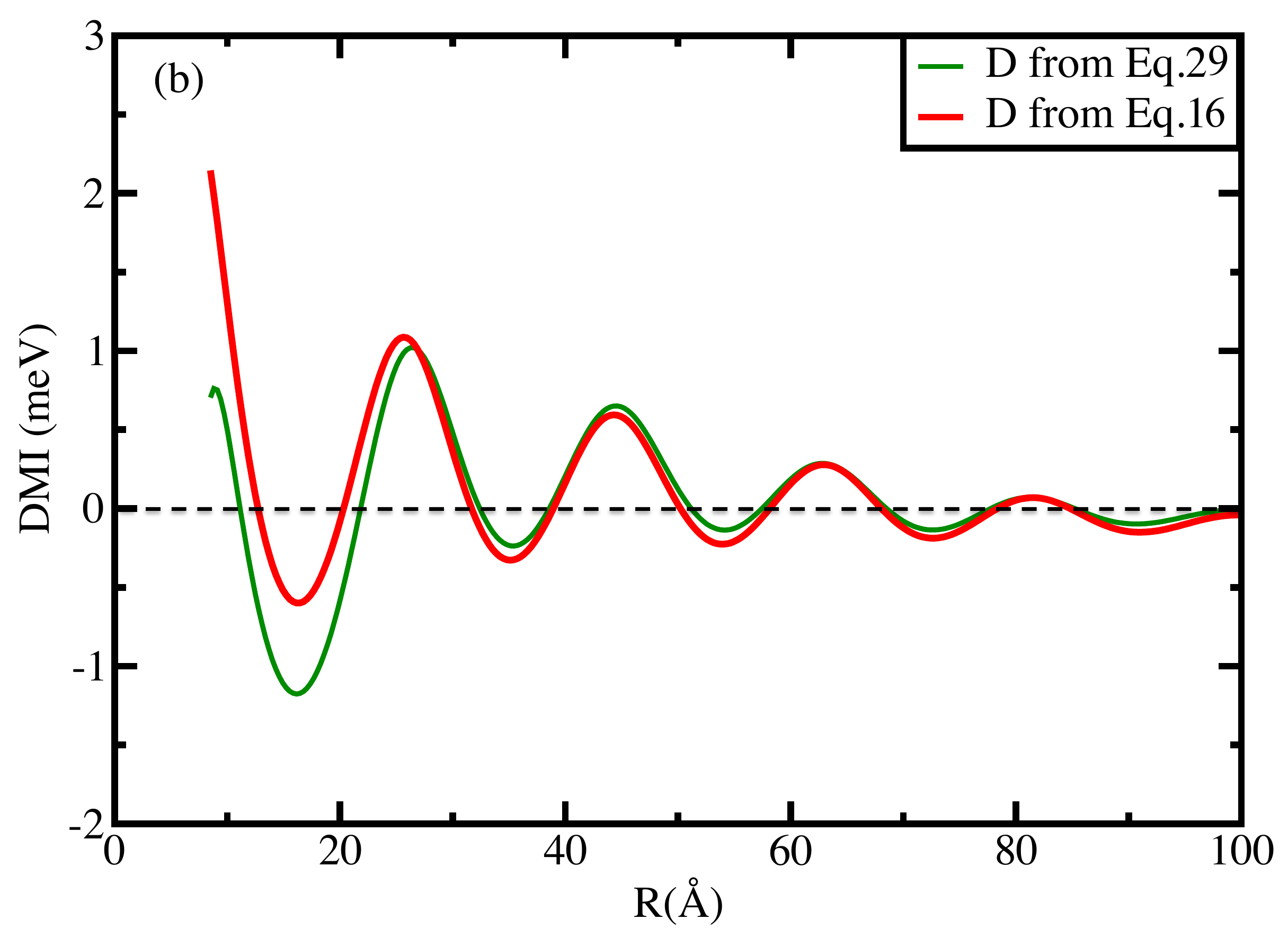}
  \caption{(a) Comparison between $D$ computed from the RKKY-approximation, Eq.~\ref{D_analytic}, and from Eq.~\ref{JtoD}. (b) The 
comparison involves $D$ computed from the renormalized Green functions Eq.~\ref{app2}, i.e. beyond the RKKY-approximation, and from Eq.~\ref{JtoD}.}
  \label{compareJtoD}
\end{figure}

Instead of the micromagnetic model, we explore in the following the possibility of relating directly $J$ and $D$. 
We noticed that the derivative of $J$ with respect to $k_{so}$ is proportional to $D$ in the RKKY-approximation:
\begin{equation}
D = \frac{1}{2R}\,\frac{\partial J}{\partial k_{\mathrm{so}}} + \sin(2k_{\mathrm{so}}R)\,\mathrm{SI}(2k_{F}R)\quad .
\label{JtoD1}
\end{equation}
Once more, if there was no sine integral we would have found a nice way of relating $D$ to $J$. Indeed, the first-order change of $J$ with 
respect to spin-orbit interaction would lead to the DM interaction $D$:
\begin{equation}
D = \frac{1}{2R}\,\frac{\partial J}{\partial k_{\mathrm{so}}}\quad .
\label{JtoD}
\end{equation}
As for the relation of Kim et al.~\cite{Kyoung-Whan}, the error is expected to be large at small distances since 
 $\mathrm{SI}(x) \sim x$. The second term in Eq.~\ref{JtoD1} cannot be neglected as shown in 
Fig.~\ref{compareJtoD}(a). While the oscillatory behavior of $D$ calculated with Eq.~\ref{JtoD} is similar to what is found from 
Eq.~\ref{D_analytic}, the magnitude of the oscillations and the sign of the interaction is very different. 
Interestingly, using the renormalized Green functions instead of the RKKY-approximation seems to ameliorate the sine integral issue, similarly 
to what was found for the magnetic interactions. We recall that in the RKKY-approximation the 
asymptotic behavior for $J$ and $D$ was peculiar since 
the sine integral led to a constant shift of the oscillations. This shift was removed when properly renormalizing the Green functions. 
In the latter case the impact of the quasi one dimensional behavior of the Rashba gas is reduced and the typical van-Hove singularity in the 
electronic density of states is decreased washing out the contribution of the sine integral. As shown in Fig.~\ref{compareJtoD}(b), utilizing 
Eq.~\ref{JtoD} leads then to a more satisfactory description of the exact result.

The intriguing implication of Eq.~\ref{JtoD} is that is gives an interpretation for the origin of the chirality being left-- or right--handed 
according to the sign of $D$. For a given distance $R$, $D$ can be of the same (opposite) sign of J if the laters's magnitude increases (decreases) 
with the spin-orbit interaction.

\section{Magnetic properties of other structures}
In this section we build magnetic nanostructrures of differents sizes and shapes made of Fe adatoms deposited 
on Au(111). We compute the magnetic interactions for the considered nanostructrures. A summary
of the obtained average magnetic interactions between nearest neighbors is provided in Table.~\ref{Exchange_table}.

\subsection{Magnetism of linear chains} 
\begin{figure}[ht]
  \centering
  \includegraphics[width= 6.5 in, clip=true]{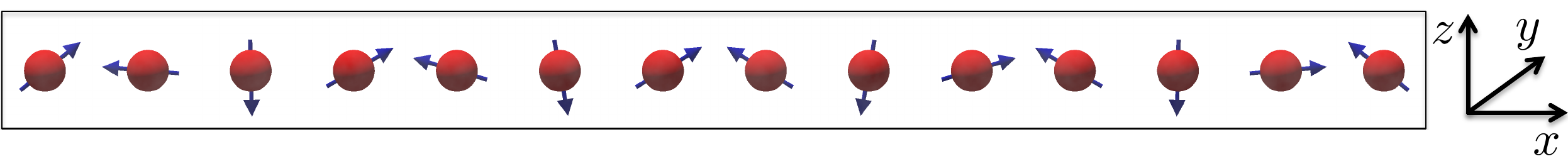}
  \caption{Magnetic ground state of a wire made of 14 adatoms. The spiral is characterized by an average rotation angle of  $110^\circ$  between 
nearest neighboring magnetic moments.}
  \label{chaineafm}
\end{figure}

Besides dimers, we investigated several linear chains of different sizes. All of them presented the same characteristics. 
Here we discuss the example of a wire made of 14 adatoms. The distance between the first nearest neighbors is chosen to 
be $d=10.42$~\AA~which corresponds to the seventh nearest neighbors distance on Au(111), where the lattice parameter $a=2.87$ \AA. 
This is very close to what is accessible experimentally\cite{Khajetoorians2}. In this case, the isotropic exchange 
interaction between the nearest-neighbors is antiferromagnetic. In average 
it is equal to $6.90$ meV, i.e. the double of the isotropic interaction obtained for the dimer, which highlights the 
impact of the nanostructure in renormalizing the electronic structure of the system. 
Within the RKKY-approximation, the magnetic interactions  would be independent from the nature, shape, size of 
the deposited nanostructures. Due to the Moriya rules, the DM vector lies along the 
$y$-direction within the surface plane similar to the dimer case. 
It is thus perpendicular to the $x$-axis defined by the chain axis. The DM interaction is around $1.99$ meV
between nearest neighbors, i.e. once more the double of the value obtained for the dimer. 
The magnetic exchange interactions are not limited to the nearest neighbors and follow an oscillatory behavior 
as function of distance. We compute the ground state starting from differents initial configurations
and compute the magnetic states where the torque acting on each magnetic moment is zero, then compare 
the energy of the obtained magnetic states and select the most stable state. It consists of a spiral
 contained in the $(xz)$ plane with an average rotation angle
 of $110^\circ$ between two nearest neighboring magnetic moments (see Fig.~\ref{chaineafm}). 
Interestingly, this angle is much smaller than the one found for the dimer ($164^\circ$) but similar to that found 
for intermediate chains sizes. The pseudo-dipolar term is around $I = 0.26$ meV, it has no 
impact on the ground state since the magnetic moments are contained in the $(xz)$ plane as aformentionned. 
This situation is equivalent to the $NC$ phase of the dimer, $I$ does not play a role determining the ground state.
 Of course, choosing an inter-atomic distance with a large pseudo-dipolar term for the dimers, 
leads generally to stable collinear magnetic wires (not shown here). We noticed that the effect of 
the magnetic anisotropy energy ($K = -6$ meV ) is mainly on the edge atoms. Indeed, the rotation angles between 
adjacent inner-moments remain around $110^\circ$ while at the edges, the magnetic moments 
are pointing more along the $z$-direction. The rotation angle between the magnetic moment at the 
edge and the $z$-axis is reduced to $25^\circ$.
\subsection{Magnetism of compact structures} 
After the one-dimensional case, we address in this section compact structures with the same interatomic distance as the one 
considered for the wire.

\begin{figure}[!h]
  \centering
  \includegraphics[width= 6.0 in, clip=true]{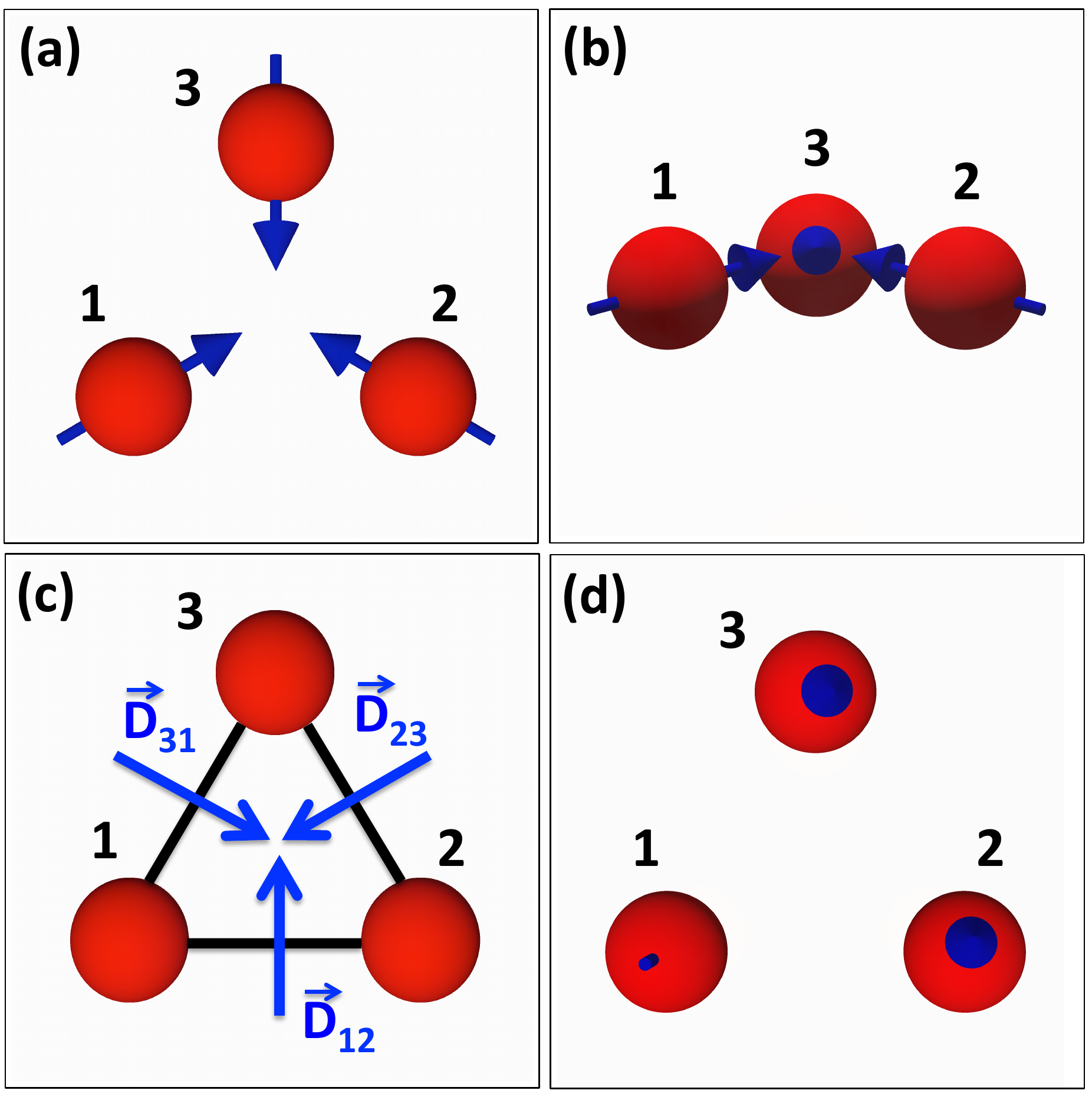}
  \caption{Top view (a) and side view (b) of the non-collinear magnetic configuration for a trimer with 
an equilateral triangle ($K = 0$ meV). While the antiferromagnetic $J$ leads to 
the $120^\circ$ configuration, the DM interaction induces a slight upward tilting of the magnetic moments. 
The corresponding DM vectors are plotted in (c). (d) Top view for the magnetic ground state of the trimer with $K = -6$ meV.}
  \label{momenttrimer1}
\end{figure}

{\bf Trimer.} We studied a trimer forming an equilateral triangle. The isotropic exchange constant $J$ is equal to $3.51$ meV favoring 
antiferromagnetic coupling, a value close to the one found for the dimer. The frustration is large in this case 
leading to a non-collinear ground state even without SO coupling\cite{Antal,Lounis_JPCM}.
 The magnetic moments lie in the same plane, e.g. the surface plane, 
with an angle of $120^\circ$ between two magnetic moments. This state has continuous degeneracy,
since rotating each magnetic moment in the same way leaves the energy invariant. 
If we now consider the DM interaction, we find that $\vec{D}$, with a magnitude of $1.0$ meV (similar to the dimer's value), 
lies in the $xy$ plane and perpendicular to the axis connecting two adatoms (see Fig.~\ref{momenttrimer1}(c)). 
This interaction lifts the degeneracy present without $D$. As depicted in Fig.~\ref{momenttrimer1}(a) and (b). 
The pseudo-dipolar term $I$ is equal to $0.13$ meV and is small compared to $J$ and $D$ therefore the non-collinear phase is more stable.
The isotropic interaction keeps the angle between the in-plane projections of the moment at $120^\circ$, while the DM interaction 
generates a slight upward tilting ($81^\circ$ instead of $90^\circ$). 
 In fact, every DM vector connecting two sites favors 
the non-collinearity of the related magnetic moments  by keeping them in the plane perpendicular to the surface and containing 
the two sites. This is however impossible to satisfy at the same time for the three pairs of atoms forming the trimer, which leads 
to the compromise shown in Fig.~\ref{momenttrimer1}(a) and (b). 
The magnetic anisotropy reduces ($K = -6$ meV) considerably the non-collinearity 
and the three moments are enforced to point almost-parallel to the $z$-axis. Two of the magnetic moments are characterized 
by an angle of $10^\circ$ instead of $81^\circ$ with respect to the $z$-axis, while the angle of the third moment is $173^\circ$ 
as shown in Fig.~\ref{momenttrimer1}(d). This is an interesting outcome compared to the behavior of the wire, 
which is characterized by a large averaged DM interaction in comparison to the trimer. Obviously the shape 
of the nanostructure is important in stabilizing non-collinear magnetism. 

\begin{table}[!h]
\centering
  \begin{tabular}{| l | l | l | l | l |}
    \hline 
    Structures & J (meV)& D (meV) & I (meV) & $\theta$($^\circ$) \\ 
    \hline
     Chain        &  6.90 & 1.99 & 0.26 & 110 \\ \hline
     Trimer       &  3.51 & 1.00 & 0.13 & 117 \\ \hline
     Hexagone  & 5.64 & 1.67 & 0.23 &  164 \\ \hline
     Heptamer  & 4.69 (4.62) & 1.37 (1.36) & 0.18 (0.12)& 120 (142) \\ \hline
  \end{tabular}
\caption{Summary of the average magnetic interactions between nearest neighbors for the calculated magnetic nanostructures.
 The values between parenthesis for the heptamer are for the nearest neighbors on the outer ring.}
\label{Exchange_table}
\end{table}

\begin{figure}[!h]
  \centering
  \includegraphics[width=6.5 in, clip=true,angle = 0]{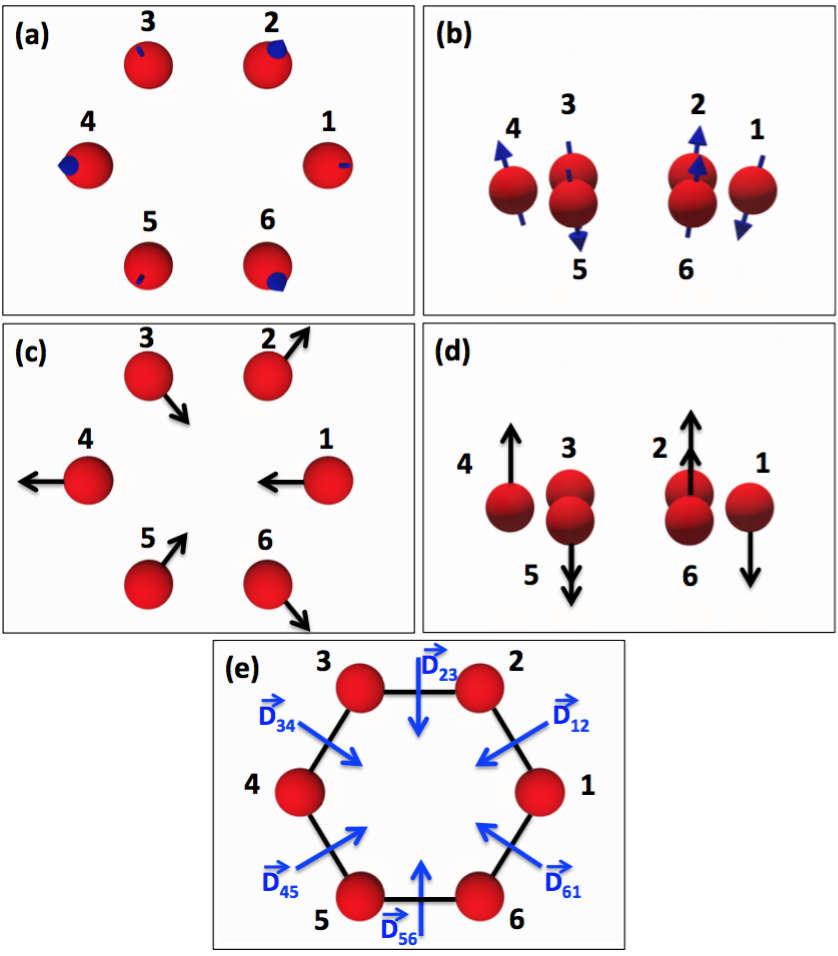}
  \caption{Top (a) and side (b) view of the magnetic ground state configuration for a hexagon made of six atoms. The projection 
of the unit vectors of the magnetic moments on the surface plane is given in (c) and the projection along the $z$-axis in (d). The 
corresponding DM vectors between the nearest neighbors are plotted in (e).}
  \label{hexagoneground}
\end{figure}

{\bf Hexagonal.} We consider now a system of six atoms forming
 a hexagonal shape with the same interatomic distance as the one considered earlier. 
The magnetic ground state configuration is non-collinear as shown in Figs.~\ref{hexagoneground}(a) and (b). 
The isotropic magnetic exchange interaction, $J$, between nearest neighbors is of antiferromagnetic type 
similarly to the value obtained for the other nanostructures studied so far. $J$ reaches a value of 
5.64 meV, which is rather close to the interaction found for the wire. In fact one could consider this hexagonal structure as 
a closed wire. The magnitude of the DM vector connecting two  nearest neighbors is large, $1.67$ meV, but not as large as the one of 
the wire. The non-collinear state is better appreciated when plotting the 
projection of the moments unit vectors on the surface plane in Fig.~\ref{hexagoneground}(c) and along the $z$-axis  
in Fig.~\ref{hexagoneground}(d). The sequence of the polar angles for every pair of nearest neighboring magnetic moments 
is given by $(16^\circ,164^\circ)$ and 
the azimuthal angle follows the symmetry of the hexagon, leading to an angle difference of $120^\circ$ between adjacent moments. 
The magnetic texture is a compromise involving the antiferromagnetic $J$ and the DM vectors 
(plotted in Fig.\ref{hexagoneground}(e)). While $J$ tries to make 
the moments anti-parallel to each other, the DM vector tends to make them lie in the plane perpendicular to the surface 
and containing at the same time the two pairs of atoms (similar to the dimer configuration). However, the magnetic moment 
has to satisfy the DM vectors arising from its nearest neighbors and therefore, the moment compromises and lies 
in the plane perpendicular to the surface and containing the atom of interest and the center of the hexagon. 
This is similar to what was found for the compact trimer. To test the stability of the non-collinear structure, we add the 
magnetic anisotropy energy and the sequence of polar angles changes from $(16^\circ, 164^\circ)$ to 
$(9^\circ, 171^\circ)$, i.e.~a change of $\approx 5^\circ$, which is not that large. The interaction
with the second and third nearest neighbors are respectively $0.40$ meV and $0.16$ meV,  
which are small compared to the first nearest neighbors. Thus, they will not affect the ground state considerably.

\begin{figure}[!h]
  \centering
  \includegraphics[width=6.5 in, clip=true,angle = 0]{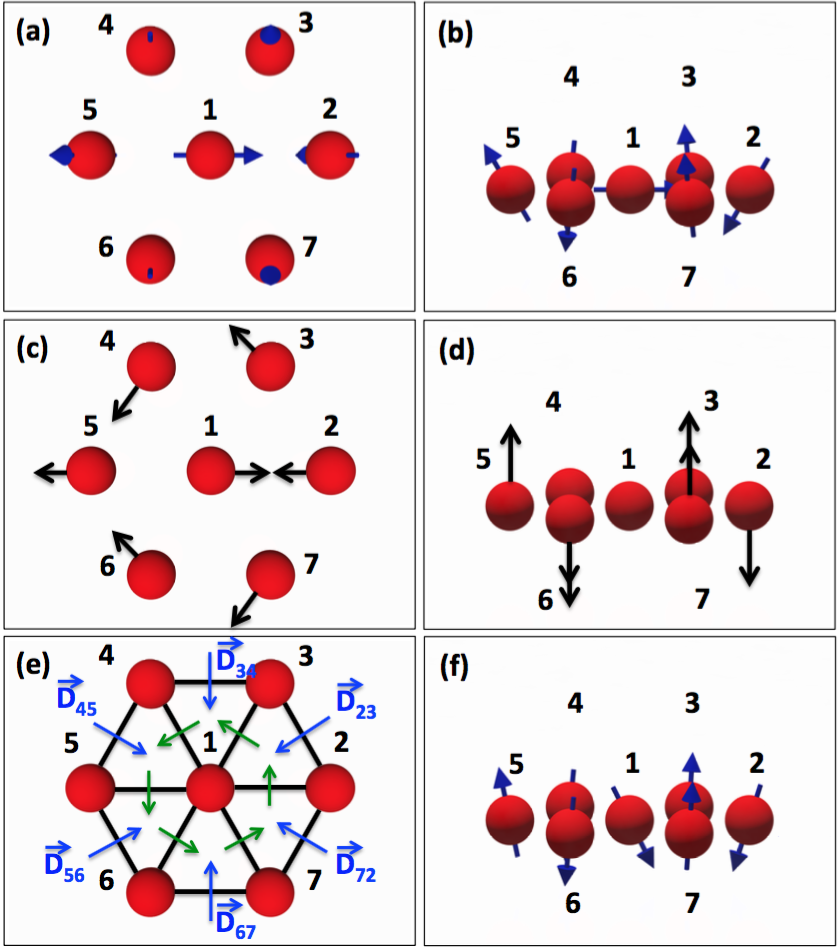}
  \caption{Top (a) and side (b) view of the magnetic ground state configuration for a hexagon made of seven atoms. The projection 
of the unit vectors of the magnetic moments on the surface plane is given in (c) and along the $z$-axis in (d). The 
corresponding DM vectors between the nearest neighbors are plotted in (e). In (f) the side view of the ground state after 
adding a single-ion magnetic anisotropy, $K$, of -6 meV.}
  \label{hexa7atomes}
\end{figure}

{\bf Heptamer.} We add to the previous structure an atom in the center of the hexagon. 
Contrary to the other atoms this central atom has six neighbors and the magnetic ground state is profoundly affected by 
this addition as shown in Fig.~\ref{hexa7atomes}(a-b). 
The nearest neighbor isotropic exchange constant $J$, 4.69 meV, decreases slightly in comparison 
to the value found for the open structure.
The obtained magnetic texture can be explained from the nearest neighboring DM interaction (1.37 meV) with the corresponding vectors
plotted in Fig.~\ref{hexa7atomes}(e). 
The addition of the central atom creates frustration similar to the trimer case. Ideally, 
every pair of nearest neighboring moments have to lie in the same plane. Thus, the central magnetic moment has to lie 
within one of the three planes orthogonal to the surface and passing by two of the outer atoms 
and the central one. In this configuration, the three atoms are satisfied and the 4 atoms left  
have the direction of their moments adjusted, which leads to the final spin-texture. 
Fig.~\ref{hexa7atomes} (c) and (d) show respectively the projection of the 
magnetic moment along the $z$-axis and in the surface plane.
Interestingly, when the single-ion magnetic anisotropy is added only the central moment is affected. 
It experiences a switch from the in-plane configuration to a quasi out-of-plane orientation. 
A side view is shown in Fig.~\ref{hexa7atomes}(f). This is another nice example 
showing how the stability of the non-collinear behavior is intimately related to the nature, shape, and size of the nanostructure.
\section{Conclusions}
We investigated the complex chiral magnetic  behavior of nanostructures of different shapes and sizes wherein the 
atoms interact via long-range interactions mediated by Rashba electrons. We use an embedding technique based on the Rashba 
Hamiltonian and the s-wave approximation followed by a mapping procedure to an extended Heisenberg model. 
The analytical forms of the elements of the tensor of the magnetic exchange interactions is presented  within the 
RKKY-approximation, i.e.~without renormalizing the electronic structure because of the presence of the nanostructure. We 
corrected the forms given by Imamura et al.\cite{Bruno2004}, and demonstrate the deep link between the magnetic interaction 
and the components of the magnetic Friedel oscillations generated by the single adatoms. The isotropic interaction and the 
DM interactions correspond respectively to the induced out-of-plane and in-plane magnetization. Besides these two interactions, the 
pseudo-dipolar term, already found in Ref.~\onlinecite{Bruno2004}, is shown to be large, generating a collinear phase competing with 
non-collinear structures induced by the DM interaction. We go beyond the RKKY-approximation by considering energy dependent scattering 
matrices and multiple scattering effects to demonstrate that the size and shape of the 
nanostructures have a strong impact on the magnitude and sign of the magnetic interactions.
We proposed an interesting connection between the DM interaction and the isotropic magnetic exchange interaction, $J$. The DM interaction 
can be related to the first order change of $J$ with respect to the spin-orbit interaction and even more important, the origin of the sign 
of the DM interaction, i.e. defining the chirality, can be interpreted by the increase or decrease of $J$ upon application of the spin-orbit interaction. 
 We considered nano-objects that can be built experimentally (see e.g. Refs.\cite{Zhu,Khajetoorians2,Khajetoorians}) and show that each of the objects 
behave differently and the stability of their non-collinear chiral spin texture is closely connected with the type of structure built 
on the substrate.
\begin{acknowledgements}
We gratefully acknowledge funding under HGF YIG Program VH-NG-717 
(Functional Nanoscale Structure and Probe Simulation Laboratory--Funsilab), the ERC Consolidator grant DYNASORE  
and the DFG project LO 1659/5-1. S.B.  acknowledges  
funding under the DFG-SPP 1666 ``Topological Insulators: Materials -- 
Fundamental Properties -- Devices''. A. Z. thanks the Algerian Ministry 
of Higher Education and Scientific Research 
for funding his sabbatical year at the 
Forschungszentrum J\"ulich.
\end{acknowledgements}
\begin{appendices}
\section{} 
\label{appendixA}
The Green function for the Rashba electron gas can be calculated using the spectral representation: 
\begin{equation}
\boldsymbol{G}^{0}(\vec{r},\vec{r}^{\,\prime},E+i\epsilon) = \sum_{\vec{k}n}\frac{\psi_{\vec{k}n}
(\vec{r})\, \psi^{*}_{\vec{k}n}(\vec{r}^{\,\prime})}{E-E_{n}+i\epsilon}
\label{spectal}
\end{equation}
Where $E_{n}$ and $\psi_{\vec{k}}(\vec{r})$ are respectively the eigenvalues and eigenstates of the Rashba Hamiltonian. 
The Rashba Green function is translationally invariant therefore $\boldsymbol{G}^{0}(\vec{r},\vec{r}^{\,\prime},E+i\epsilon) 
= \boldsymbol{G}^{0}(\vec{R},E+i\epsilon)$, with $\vec{R} = \vec{r}-\vec{r}^{\,\prime}$. After performing the sums over $\vec{k}$ and $n$, 
the diagonal and off diagonal spin elements of the Green function $\boldsymbol{G}^{0}$ of the Rashba electrons are given as: 
\begin{equation}
G_D(R,E+i\epsilon) = -\frac{im^*}{2\hbar^2(k_1+k_2)}\Bigg{[}k_1\,H_{0}(k_1R + i\epsilon)+k_2\,H_0(k_2R + i\epsilon)\Bigg{]}\quad ,
\label{g01}
\end{equation}
\begin{equation}
G_{ND}(R,E+i\epsilon) = -\frac{im^*}{2\hbar^2(k_1+k_2)}\Bigg{[}k_1\,H_1(k_1R + i\epsilon)-k_2\,H_1(k_2R + i\epsilon)\Bigg{]}\quad .
\label{g02}
\end{equation}
As mentioned in the main text, the vectors $k_{1}$ and $k_{2}$ are given by $k_1 = k_\mathrm{so} + 
\sqrt{k_\mathrm{so}^2+\frac{2m^*E}{\hbar^2}}$ and $k_2=- k_\mathrm{so} + \sqrt{k_\mathrm{so}^2+\frac{2m^*E}{\hbar^2}}$ with $k_\mathrm{so} 
= \frac{m^* \alpha}{\hbar^2}$.\\
\section{} 
\label{appendixB}
In this appendix we derive the generalized Heisenberg Hamiltonian $H_{ij} = \vec{e}_{i}\,\doubleunderline
{{\mathbf{J}}}_{ij}\,\vec{e}_{j}$, which was simplified to the form given by Eq.~\ref{zest}. For this purpose, we 
need to calculate the elements of the tensor of exchange interactions showing    
up in Eq.~\ref{app2}, i.e. $\mathrm{Tr}\,\{\boldsymbol{\sigma}^\alpha\,\mathbf{G}_{ij}\,
\boldsymbol{\sigma}^\beta\,\mathbf{G}_{ji}\}$, considering that $\mathbf{G}$ can be expressed in terms of 
$G_{D}$ and $G_{ND}$ (see Eq.~\ref{G_pauli}). This can be evaluated via the following trace (omitting the energy integration):
\begin{equation}
\begin{split}
H_{ij} & = \mathrm{Tr}\,[(\vec{e}_i \cdot \vec{\boldsymbol{\sigma}})(G_{D}\,\boldsymbol{\sigma}_{0} 
- i\,G_{ND}(\cos\beta\,\boldsymbol{\sigma}_y - \sin\beta\,\boldsymbol{\sigma}_x))\\
&\hspace{10mm}\times(\vec{e}_j\cdot\vec{\boldsymbol{\sigma}})(G_{D}\,\boldsymbol{\sigma}_{0} + i\,G_{ND}(\cos\beta\ 
\boldsymbol{\sigma}_y - \sin\beta \hspace{1mm}\boldsymbol{\sigma}_x))]\quad .
\end{split}
\end{equation}
Using the properties of the Pauli matrices, we know that for two vectors $\vec{A}$ and $\vec{B}$, the following relation holds: 
$(\vec{A}\cdot\vec{\boldsymbol{\sigma}})\,(\vec{B}\cdot\vec{\boldsymbol{\sigma}}) = (\vec{A}\cdot\vec{B})\ 
\boldsymbol{\sigma}_{0} + i\,(\vec{A}\times\vec{B})\cdot\vec{\boldsymbol{\sigma}}$. Thus :
\begin{equation}
\begin{split}
H_{ij}& =  2\,\vec{e}_i\cdot\vec{e}_j\,(G_{D}^2 - G_{ND}^2) - 4\,(\vec{e}_i\times\vec{e}_j)_x\,i\,G_D\,G_{ND}\,\sin\beta\\
& \quad- 4\,(\vec{e}_i\times\vec{e}_j)_y\,G_{D}\,G_{ND}\,\cos\beta + 4\,e_i^y\,e_j^y\,G_{ND}^2\,\cos^2\beta\\
& \quad+4\,e_i^x\,e_j^x\,G_{ND}^2\,\sin^2\beta - 2\,(e_i^x\,e_j^y + e_i^y\,e_j^x)\,G_{ND}^2\,\sin\beta\,\cos\beta\quad . 
\end{split}
\end{equation}
The terms proportional to $e_i^x\,e_j^x$ and $e_i^y\,e_j^y$ will lead to the pseudo-dipolar like terms after performing the energy integration 
given in Eq.~\ref{app2}. The terms proportional 
to $(e_i^x\,e_j^y + e_i^ y\,e_j^x)$ are called interface terms. We can combine both terms in a pseudo-dipolar Hamiltonian for the two-dimensional 
case;
\begin{equation}
H_{psd} = I\,\sum_{i,j} [(\vec{e}_i\cdot\vec{e}_j) - (\vec{e}_i\cdot\vec{R}_{ij})  (\vec{e}_j\cdot\vec{R}_{ij}) - e_{i}^z\,e_{j}^z]\quad .
\end{equation}
$\vec{R}_{ij}$ is the vector connecting the impurities \{i, j\}. 

If we consider that the two magnetic 
impurities are along the $x$-axis then $\beta = 0$ and we get the expression below for the trace: 
\begin{equation}
H_{ij} =  2\,(G_{D}^2 - G_{ND}^2)\,\vec{e}_i\cdot\vec{e}_j - 4\,G_{D}
\,G_{ND}\,(\vec{e}_i\times\vec{e}_j)_y + 4\,G_{ND}^2 e_i^y\,e_j^y\quad , 
\end{equation} 
which leads to the final form of the Hamiltonian given in Eq.~\ref{zest}, and to the identification of the different magnetic interaction 
terms as presented in Eqs.~\ref{J_analytic}, \ref{D_analytic}, \ref{I_analytic}.
\section{} 
\label{appendixC}
In order to obtain the analytical forms of $J$, $D$ and $I$ in the RKKY-approximation 
(Eqs.~\ref{J_analytic}, \ref{D_analytic}, \ref{I_analytic}), we evaluate the integrands 
needed in Eqs.~\ref{ech1}, \ref{ech2}, \ref{ech3} considering two regimes, positive or negative $k_1$. For $k_1 < 0$:
\begin{equation}
G_{D}^2 = -\frac{(m^{*})^2}{4\hbar^2(k_1+k_2)^2}\hspace{1mm}[\hspace{1mm}k_{1}^2\hspace{1mm}
H_0^{*2}(|k_1|R)+k_{2}^2\hspace{1mm}H_{0}^{2}(k_2R)-2\ k_1k_2\hspace{1mm}H^{*}_{0}(|k_1|R)H_{0}(k_2R)\hspace{1mm}]\quad ,
\end{equation}
\begin{equation}
G_{ND}^2 =
-\frac{(m^{*})^2}{4\hbar^2(k_1+k_2)^2}\hspace{1mm}[\hspace{1mm}k_{1}^2\hspace{1mm}
H_1^{*2}(|k_1|R)+k_{2}^2\hspace{1mm}H_{1}^{2}(k_2R)-2\ k_1k_2\hspace{1mm}H^*_1(|k_1|R)H_{1}(k_2R)\hspace{1mm}]\quad ,
\end{equation}
and 
\begin{equation}
\begin{split}
G_{D}G_{ND}&= - \frac{(m^{*})^2}{4\hbar^2(k_1+k_2)^2}\hspace{1mm}[\hspace{1mm}-k^{2}_{1}
\hspace{1mm}H^{*}_{0}(|k_1|R)H^{*}_{1}(|k_1|R)+k_{1}k_{2}\ H_{0}^{*}(|k_1|R)H_{1}(k_2R)\\
& \hspace{25mm} \quad +k_{1}k_{2}\ H_{1}^{*}(|k_1|R)H_{0}(k_2R)-k_2^2\hspace{1mm}H_{1}(k_2R)H_{0}(k_2R)\hspace{1mm}]\quad .
\end{split}
\end{equation}
In case $k_1 > 0$:
\begin{equation}
G_{D}^2 = -\frac{(m^{*})^2}{4\hbar^2(k_1+k_2)^2}\hspace{1mm}[\hspace{1mm}k_{1}^2
\hspace{1mm}H_{0}^2(k_1R)+k_{2}^2\hspace{1mm}H_{0}^2(k_2R)+2k_1k_2\hspace{1mm}
H_0(k_1R)H_0(k_2R)\hspace{1mm}]\quad ,
\end{equation}
\begin{equation}
G_{ND}^2 = -\frac{(m^{*})^2}{4\hbar^2(k_1+k_2)^2}\hspace{1mm}[\hspace{1mm}k_{1}^2
\hspace{1mm}H_{1}^2(k_1R)+k_{2}^2\hspace{1mm}H_1^2(k_2R)-2k_1k_2\hspace{1mm}
H_1(k_1R)H_1(k_2R)\hspace{1mm}]\quad ,
\end{equation}
and 
\begin{equation}
\begin{split}
G_{D}\ G_{ND}&= - \frac{(m^{*})^2}{4\hbar^2(k_1+k_2)^2}\hspace{1mm}[\hspace{1mm}k_1^2
\hspace{1mm}H_{0}(k_1R)H_{1}(k_1R)-k_1k_2\hspace{1mm}H_{0}(k_1R)H_{1}(k_2R)\\
& \hspace{25mm} \quad +k_1k_2\hspace{1mm}H_{1}(k_1R)H_{0}(k_2R)-k_2^2\hspace{1mm}H_{0}(k_2R)H_{1}(k_2R)\hspace{1mm}]\quad .
\end{split}
\end{equation}
We use the asymptotic expansion for the Hankel functions for large R: $H_0(x)\simeq \sqrt{\frac{2}{\pi x}}\hspace{1mm} e^{i(x-\frac{\pi}{4})}$ and 
$H_1(x)\simeq \sqrt{\frac{2}{\pi x}}\hspace{1mm} e^{i(x-\frac{3\pi}{4})}$ which simplify the previous forms for negative $k_1 < 0$ to:
\begin{equation}
G_{D}^2 = \frac{i(m^{*})^2}{2\hbar^2(k_1+k_2)^2\pi R}\,[-|k_1|\, 
e^{-2i|k_1|R} + k_2\,e^{2ik_2R} + 2i\,\sqrt{|k_1| k_2}\,e^{(k_2-|k_{1}|) R}\,]\quad ,
\end{equation}
\begin{equation}
G_{ND}^2 = -\frac{i(m^{*})^2}{2\hbar^2(k_1+k_2)^2\pi R}\,[-|k_1|\, 
e^{-2i|k_1|R} + k_2\,e^{2ik_2R} - 2i\,\sqrt{|k_1|k_2}\,e^{(k_2-|k_{1}|) R}\,]\quad ,
\end{equation}
and
\begin{equation}
G_{D}\,G_{ND} = \frac{(m^{*})^2}{2\hbar^2(k_1+k_2)^2\pi R}\,[\,-|k_1| 
\ e^{-2i|k_1|R} - k_2\,e^{2i k_2 R}\,]\quad .
\end{equation}
While a positive $k_1$ leads to:
\begin{equation}
G_{D}^2 = \frac{i(m^{*})^2}{2\hbar^2(k_1+k_2)^2\pi R}\,[\,k_1\,e^{2ik_1R} +  k_2\,
e^{2ik_2R} + 2\sqrt{k_1k_2}\,e^{i(k_1+k_2)R}\,]\quad ,
\end{equation}
\begin{equation}
G_{ND}^2 = -\frac{i(m^{*})^2}{2\hbar^2(k_1+k_2)^2\pi R}\,[\,k_1 
\,e^{2ik_1R} +  k_2\,e^{2ik_2R} - 2\sqrt{k_1k_2}\,e^{i(k_1+k_2)R}\,]\quad ,
\end{equation}
and
\begin{equation}
G_{D}\,G_{ND} = \frac{(m^{*})^2}{2\hbar^2(k_1+k_2)^2\pi R}\,[\,k_1 
\,e^{2ik_1R} - k_2\,e^{2ik_2R}\,]\quad .
\end{equation} 
From the expressions above we notice that contrary to the terms $(G_{D}^2 -G_{ND}^2)$ and $G_{D}\ G_{ND}$,  $G_{D}$ and $G_{ND}$ behave differently 
in the first and second regime.
\end{appendices}
\bibliography{bibliography}
\end{document}